\documentstyle[prb,aps]{revtex}

\begin{document}
\draft
\title{Microscopic Theory for Emission of Elementary Excitations into He II
from a Heated Solid.}
\author{Alexander V. Zhukov}
\address{Institute of Physics, Academia Sinica, Nankang,
Taipei 11529, Taiwan}
\date{\today}
\maketitle

\begin{abstract}
I develop here the microscopic quantum theory for description of
creation of phonons and rotons in superfluid helium by a solid
heater. Starting with correct transfer Hamiltonian describing a
coupling between the solid and liquid $^4$He the probabilities of
transformation of a single phonon in the solid into i) single
helium phonon, ii) two helium phonons, and iii) single helium
roton are found out. All the obtained expressions account for
different polarizations of phonons in the solid. The heat transfer
associated with single phonon and single roton channels are
calculated. Particularly, the obtained expression for heat flux
via the single phonon channel calculated in the framework of
present microscopic theory exactly coincides with the well known
Khalatnikov formula obtained initially in the framework of
acoustic-mismatch theory. The impossibility of direct creation of
R$^{(-)}$ rotons becomes clear in the used framework due to
accurate account of the boundary conditions at the solid -- liquid
helium interface, which is in agreement with recent experimental
results.
\end{abstract}

\pacs{67.40.-w, 67.40.Pm, 68.08.-p, 05.30.Jp}

\section{Introduction}

The elementary excitations in superfluid $^4$He have been the
object of extensive study for many years. The phonon--roton type
excitation spectrum proposed phenomenologycally by Landau \cite{1}
allows to account correctly for superfluid properties and
thermodynamic behaviour of liquid helium below the
$\lambda$-transition temperature. The existence of such a spectrum
was demonstrated by the numerous direct observations of single
excitations from inelastic neutron scattering. In spite of many
various attempts to built a completely microscopic theory to
derive directly the thermal excitations spectrum, the general
picture drawn first by Landau \cite{1} and Feynman and Cohen
\cite{2} had not changed essentially during more than four
decades.

Since we realized the general picture of elementary thermal
excitations in bulk liquid helium, the natural problem appeared:
how to describe a heat flow across a boundary between a solid and
liquid helium. All earlier investigations of this process were
concentrated on studying of a discontinuity in temperature
appeared between the two materials \cite{3}, which is commonly
known now as Kapitza resistance. The first consistent accurate
theory of this effect was developed by Khalatnikov \cite{4}. He
built his theory starting with quantizing of elastic waves of
different polarizations in the solid. But in general, the main
results of Khalatnikov's theory are completely consistent with
classical acoustic--mismatch theory due to Little \cite{5}. The
in-depth reviews of theoretical and experimental investigations of
Kapitza resistance problem are given by Challis \cite{6} and
recently by Zinov'eva \cite{7}. In their pure form all these
theories show that transfer of phonons from the solid to the
liquid helium is impeded first due to a large difference in
densities $\rho_L/\rho_S\ll 1$ ($\rho_L$ and $\rho_S$ is the
densities of liquid helium and solid, respectively), and second
due to the smallness of critical angle $\sin\theta_c= c/s_t\ll 1$,
where $c$ is the first sound velocity in helium and $s_t$ is the
velocity of transverse sound in solid. The total internal
reflection occurs when the angle of incidence exceeds this angle
$\theta_c$. Because of this the helium phonon can propagate only
into a narrow cone along the direction perpendicular to the
interface.

 \par
According to the present understanding, solid can inject the
phonons in two groups \cite{8}. The first is phonons in the narrow
cone perpendicular to the solid heater surface, while the second
group is phonons emitted into the entire half space, these are the
so-called background phonons. The existence of the first group is
understood satisfactory from acoustic--mismatch theories
\cite{4,5,6}. This mechanism was directly confirmed by experiments
\cite{9} (see also Ref. 7). However the existence of background
phonons, which is evident from experiments on phonon emission from
cleaved crystals \cite{9,10}, was not satisfactory explained
theoretically until now in spite of many attempts \cite{11}. In
fact, all these theories were devoted to various modifications of
the acoustic--mismatch theory to account the imperfection of the
interface and some additional scattering mechanisms. As a result,
only small corrections to the Khalatnikov theory were obtained.

 \par
All theoretical works cited above have been based on the classical
acoustic theory and there was thus need for a microscopic quantum
mechanical description since $^4$He is a quantum liquid rather
than a classical elastic fluid. There were several attempts to
build a quantum mechanical theory of solid--liquid helium heat
transmission based on various simplified models \cite{12}, but the
first really essential contribution to the subject has been done
by Sheard, Bowley, and Toombs \cite{13} (hereinafter is referred
as SBT). In fact I present here the modification of this theory.

 \par
The SBT theory is based on the description of coupling between the
solid and the liquid $^4$He via a transfer Hamiltonian in analogy
with the theory of electron tunneling through a
metal-insulator-metal junction. The main insight of Ref. 13
consist in the expression of tunnel Hamiltonian through the
displacement of the solid at the interface and the stress tensor
of liquid helium.

 \par
The main shortcoming of SBT theory is the clear internal asymmetry
of their transfer Hamiltonian with respect to solid and He II.
Namely, as it follows from their theory, the solid acts on the
helium, but there is no return action. As a result the transfer
coefficient for the single phonon channel (single solid phonon --
single helium phonon) differs from the result obtained by
Khalatnikov \cite{4}, which is correct for this process without
any doubt. The authors of Ref. 13 have also roughly estimated the
transfer coefficients for single solid phonon -- two helium
phonons (two phonons process) and single solid phonon -- single
roton processes. As it will be shown below the more accurate
computation of these coefficients gives rather different results.

 \par
Calculation of probability for creation of rotons by a solid
heater is called for by recent experiments of Exeter group
\cite{14}, which obtained quite a number of very interesting
results. Particularly, they concluded that creation of the
R$^{(+)}$ rotons (rotons with wave vector greater than
corresponding to roton minimum $k_0\approx 1.92$\AA) does not
essentially depend on the particular geometry of the heater, while
the R$^{(-)}$ rotons (rotons with wave vector smaller than
corresponding to roton minimum $k_0\approx 1.92$\AA) cannot be
created by a simple flat heater at all. This observation still
requires the proper explanation.

 \par
I this paper I propose a microscopic quantum theory for creation
of thermal excitations in superfluid $^4$He by a heated solid. The
starting point of the developed theory is the construction of more
correct transfer Hamiltonian to exploit further the main idea of
SBT theory \cite{3}. To test the developed theory I first consider
the heat flow via the single phonon channel with account of all
phonon polarizations in a solid. The obtained expression for a
heat flux through the interface associated with this process
coincides exactly with the well known Khalatnikov result \cite{4}.
This direct support of the formalism allowed me to consider also
the single roton emission process as well as to write down the
probabilities for emission of a phonon pair by a single phonon in
the solid. Particularly, I argue that emission of R$^{(-)}$ rotons
is forbidden by the boundary conditions at the solid -- He II
interface, which is in agreement with experimental observations
\cite{14}.

 \par
The paper is constructed as follows. The transfer Hamiltonian
corresponding to solid -- liquid helium interface is constructed
in Section II. In Section III I write down the tunnel Hamiltonians
for the particular single excitation and double excitations
processes. The well-known Khalatnikov formula for a transfer of
single solid phonon into single phonon in helium is obtained from
the general formalism in Section IV. Finally in Section V I study
the creation of rotons from a heated solid via the single
excitation channel. The outlines and conclusions are given in
Section VI.

\section{Solid--Liquid coupling Hamiltonian}

The whole system under consideration consists of the solid and the
superfluid helium with a common boundary between them. To study
the excitations transmission processes through the interface we
must pick up the part of total Hamiltonian, which corresponds to
the interface itself. Just this interface Hamiltonian represents
the coupling between two media.

 \par
Let $\hat\psi^\dag ({\bf r})$ and $\hat\psi ({\bf r})$ be the
field operators (creation and annihilation operators in
Scr{\"o}dinger representation) in helium. Here ${\bf r}$ is the
coordinate of Lagrange point in superfluid helium. Suppose that
thermal perturbation lead to the displacement of some atom in
solid with equilibrium coordinate ${\bf R}_n$ by vector ${\bf
u}_n$. In the frame of pair interaction approximation we can write
the deviation of a total Hamiltonian from the equilibrium one in
the following form:
 \begin{equation}
 \hat{\cal H}_{\rm{int}} = \int\limits_{\Omega_L}
 \hat\psi^\dag ({\bf r})
 \sum_n \hat{U}_{\rm{int}}
 \left(
 |{\bf R}_n+{\bf u}_n-{\bf r}-\delta{\bf r}|
 \right)
 \hat\psi ({\bf r}) d^3r,
 \label{1}
 \end{equation}
where $\Omega_L$ is the volume of the helium, $\hat{U}_{\rm{int}}$
is the potential of interaction between a physical point of helium
with a volume element in solid. This potential behaves like a
single-particle operator when act on the creation and annihilation
operators in helium. The quantity $\delta{\bf r}$ in eq. (\ref{1})
stands for the physical point displacement in helium due to the
perturbation ${\bf u}_n$ in solid. It is connected with the
density fluctuations in helium by the relation
 \begin{equation}
 \delta n({\bf r}) = n({\bf r}) - n_0 =
 \frac{\partial n}{\partial{\bf r}}\delta {\bf r}.
 \label{2}
 \end{equation}
Let us assume the interface to be a plane perpendicular to $z$th
direction as is shown in Fig. \ref{f1}. Supposing the
displacements ${\bf u}$ and $\delta{\bf r}$ to be small, we can
rewrite expression (\ref{1}) up to the first order in these
perturbations as
 \begin{equation}
 \hat{\cal H}_{\rm{int}} = \int\limits_{\Omega_L}
 d^3rn({\bf r})U_0({\bf r})+ \int\limits_{\Omega_L} d^3r
 u_z(x,y,0) f_z^{(SL)}({\bf r})- \int\limits_{\Omega_L}
 d^3r \delta r_z(x,y,0) f_z^{(LS)}({\bf r}),
 \label{3}
 \end{equation}
where
 \begin{eqnarray}
 {\bf f}^{(SL)}({\bf r})=\hat\psi^\dag ({\bf r})
 \sum_n \nabla_{{\bf R}_n}{\hat U}_{\rm{int}}
 ({\bf R}_n-{\bf r})\hat\psi ({\bf r}),
 \label{4}\\
 {\bf f}^{(LS)}({\bf r})=\hat\psi^\dag ({\bf r})
 \nabla_{\bf r}\sum_n{\hat U}_{\rm{int}}
 ({\bf R}_n-{\bf r})\hat\psi ({\bf r})
 \label{5}
 \end{eqnarray}
are the forces per unit volume of liquid helium and solid,
respectively, and
 \[
 U_0({\bf r})=\sum_n {\hat U}_{\rm{int}}({\bf R}_n-{\bf r}).
 \]
In the expression (\ref{3}) we take the quantities $u_z(x,y,0)$
and $\delta r_z(x,y,0)$ at the plane $z=Z=0$ because of assumed
short--range interaction. The first term in eq. (\ref{3})
corresponds to a static interaction and does not depend on the
local spatial perturbations. Thus, it naturally should be included
to the equilibrium Hamiltonian. To transform remaining two terms
we have to account correctly the boundary conditions at the
interface. First, require the normal components of displacements
in solid and liquid helium at the surface to be equal
 \begin{equation}
 u_z(x,y,0)=\delta r_z(x,y,0).
 \label{6}
 \end{equation}
Second, the total work done must be zero. This requirement can be
expressed mathematically as
 \begin{equation}
 \int dz f_z^{(SL)}(x,y,z) + \int dZ f_z^{(LS)}(X,Y,Z) =0.
 \label{7}
 \end{equation}
Using the relations (\ref{6}) and (\ref{7}) we can express the
interaction Hamiltonian (\ref{3}) in the following simple form
 \begin{equation}
 \hat{\cal H}_{\rm{int}}=2\int\limits_{\Omega_L} d^3r
 u_z(x,y,0) f_z^{(SL)}({\bf r}).
 \label{8}
 \end{equation}
Note, the interaction Hamiltonian (\ref{8}) obtained here differs
from SBT Hamiltonian \cite{13} by the factor $2$. This difference
follows from incorrect account of mutual influence of solid and
liquid helium by the authors of Ref. 13. They have accounted only
the perturbation in the solid, which leads to an incident
excitation. But as it can be seen from the expression (\ref{1})
above, the account of corresponding perturbation in helium leads
to a return response, which doubles the interaction Hamiltonian in
view of the boundary conditions (\ref{6}) and (\ref{7}).

 \par
Using the relation
 \[
 \nabla_{{\bf R}_n}{\hat U}_{\rm{int}}({\bf R}_n-{\bf r})=
 -\nabla_{\bf r}{\hat U}_{\rm{int}}({\bf r}-{\bf R}_n)
 \]
and the expression (\ref{4}) for ${\bf f}^{(SL)}$ we can get the
following relation
 \begin{equation}
 f_z^{(SL)}(x,y,z)\approx -n({\bf r})
 \frac{dU_0}{dz},
 \label{9}
 \end{equation}
where $n({\bf r})$ is the number density of helium atoms. Now we
come to the first essential approximation. Let us choose $U_0(z)$
in the form of repulsive potential step at the solid--liquid
interface
 \begin{equation}
 U_0(z)=
 \left\{
 \begin{array}{lr}
 V_0 & (z<0)\\
 0 & (z>0)
 \end{array}
 \right. ,
 \label{10}
 \end{equation}
where $V_0=\rm{const}>0$. Below I will use the method of
elimination of the force (\ref{9}) from Hamiltonian (\ref{8})
proposed by SBT \cite{13}. Due to their insight the height $V_0$
of the step (\ref{10}) is not crucial. Note, the combination
$-n({\bf r}) (dU_0/dz)$ is the external force density exerted on
the liquid helium by the solid. The hydrodynamic equation of
motion looks
 \begin{equation}
 \frac{\partial j_k}{\partial t}+
 \frac{\partial\pi_{kl}}{\partial x_l} =f_k^{(SL)},
 \label{11}
 \end{equation}
where ${\bf j}$ is the momentum density of liquid helium,
$\pi_{kl}$ is the fluid stress tensor. On the other hand the
quantum mechanical equation for evolution of operator ${\bf j}$
(Heisenberg equation) gives
 \begin{equation}
 \frac{i}{\hbar}\left[ j_z, \hat{\cal H}_L\right]=
 n({\bf r})\frac{dU_0}{dz}+
 \frac{\partial\pi_{zl}}{\partial x_l},
 \label{12}
 \end{equation}
where $\hat{\cal H}_L$ is the Hamiltonian of liquid helium.
Following Ref. 13 let us integrate equation (\ref{12}) in the
interval $z\in \{-\varepsilon,+\varepsilon\}$, where $\varepsilon$
is of the order of distance over which an average helium density
decays from its bulk value to zero inside the potential step. In
fact, I will put finally $\varepsilon \to 0$, which is consistent
with our choice of the potential (\ref{10}). Since the momentum
density ${\bf j}$ remains finite the integral over the left hand
side of eq. (\ref{12}) behaves like $o(z)$ and therefore can be
neglected. The derivatives of all components of vector $\pi_{zl}$
except $\partial\pi_{zz}/\partial z$ are finite as well. So, we
get
 \begin{equation}
 -\int\limits_{-\varepsilon}^{+\varepsilon}n({\bf r})
 \frac{dU_0}{dz}dz=\pi_{zz}\Bigl\vert_{z=\varepsilon}-
 \pi_{zz}\Bigr\vert_{z=-\varepsilon}=\pi_{zz}(x,y,0).
 \label{13}
 \end{equation}
Finally, using the formulae (\ref{8}), (\ref{9}), and (\ref{13})
we can write the interaction Hamiltonian in the form
 \begin{equation}
 \hat{\cal H}_{\rm{int}}=2\int\limits_{\Sigma}
 u_z(x,y,0)\pi_{zz}(x,y,0)dxdy,
 \label{14}
 \end{equation}
where $\Sigma$ is the area of solid-liquid interface. Formula
(\ref{14}) represents the interaction Hamiltonian, expressed
through the solid displacement at the surface and diagonal
component of liquid stress tensor. Now it should be represented in
terms of collective variables, which allow to carry out second
quantization in the simplest way. It is naturally to present
$\pi_{zz}$ as a sum of kinetic and potential parts as follows
 \begin{equation}
 \pi_{zz}=\pi_{zz}^{(\rm{kin})}+\pi_{zz}^{(\rm{pot})}.
 \label{15}
 \end{equation}
The explicit forms of $\pi_{zz}^{(\rm{kin})}$ and
$\pi_{zz}^{(\rm{pot})}$ are derived in Ref. 13 using the
semi--microscopic theory of superfluids developed by Sunakawa {\it
et. al.} \cite{15}. Thus, in our notations the kinetic part of
(\ref{15}) takes the form
 \begin{equation}
 \pi_{zz}^{(\rm{kin})} = m n v_z^2+
 \frac{\hbar^2}{2m}\left(
 \frac{\partial n}{\partial z}n^{-1}
 \frac{\partial n}{\partial z} -
 \frac{\partial^2 n}{\partial z^2}\right) -
 \sum_{\bf k}\frac{\hbar^2 k_z^2}{2m\Omega_L},
 \label{16}
 \end{equation}
where $m$ is the mass of helium atom, $n$ is the number density
operator, which stands for the first collective variable, $v_z$ is
$z$th component of the velocity operator, which stands for the
second collective variable, ${\bf k}$ is the wave vector of
excitation in liquid helium.

 \par
The potential part of (\ref{15}) can be obtained from the equation
of motion, which leads to the relation
 \begin{equation}
 \sum_k\frac{\partial\pi_{zz}^{(\rm{pot})}}
 {\partial x_k}= \hat\psi^\dag ({\bf r})
 \left\{
 \int d^3r' \hat\psi^\dag ({\bf r}')V({\bf r}-{\bf r}')
 \hat\psi ({\bf r}')
 \right\}
 \hat\psi ({\bf r})
 \label{17},
 \end{equation}
where $V({\bf r}-{\bf r}')$ is the inter-atomic potential in
helium. In fact the equation (\ref{17}) implies the pair
interaction, which strictly speaking is inappropriate for the
liquid helium. We however will eliminate it from all final
relations by the use of semi-phenomenological notations associated
with superfluid helium. The detailed derivation of
$\pi_{zz}^{({\rm pot})}$ from eq. (\ref{17}) can be found in Ref.
13.

 \par
In terms of Fourier transforms defined by ${\bf v}=\sum_{\bf k}
{\bf v}_{\bf k}\exp (i{\bf k}{\bf r})$ and $n=\sum_{\bf k} n_{\bf
k}\exp (-i{\bf k}{\bf r})$ the kinetic and potential parts of
$z$th diagonal element of the stress tensor up to the second order
in collective variables look
 \begin{eqnarray}
 \pi_{zz}^{(\rm{kin})}=\sum_{{\bf k},{\bf k}'}\left\{
 \rho_Lv_{-{\bf k}z}v_{-{\bf k}'z}-
 \frac{\hbar^2k_zk'_z}{4\rho_L}
 n_{{\bf k}z}n_{{\bf k}'z}
 \right\}{\rm e}^{-i({\bf k} +{\bf k}'){\bf r}}
 \nonumber \\
 +\frac{\hbar^2}{4m}\sum_{\bf k}k_z^2 n_{\bf k}
 {\rm e}^{-i{\bf k}{\bf r}}-\sum_{\bf k}
 \frac{\hbar^2 k_z^2}{2m\Omega_L},
 \label{18}
 \end{eqnarray}

 \begin{equation}
 \pi_{zz}^{(\rm{pot})}=\sum_{{\bf k},{\bf p}
 \atop{{\bf p}\neq0}}
 n_{{\bf p}-{\bf k}}n_{\bf k}V({\bf k})\frac{k_z}{p_z}
 {\rm e}^{-i{\bf p}{\bf r}}
 \label{19}
 \end{equation}
where $V({\bf k})$ is the Fourier transform of interatomic
interaction potential in helium, which will be expressed below in
terms of excitation spectrum of superfluid helium. Above we have
used only two essential approximations: the step shape of
potential (\ref{10}) and the pair interaction between helium atoms
which results in the expression (\ref{19}). The first
approximation can be naturally justified in view of smallness of
boundary inhomogeneity domain, which is of the order of
interatomic distances \cite{16}. To avoid the explicit use of the
second approximation I will act as follows. The most general
approach to the problem of low lying energy levels in superfluids
lead to the following form of excitation spectrum \cite{17}
 \begin{equation}
 \hbar\omega_{\bf k}=\frac{\hbar^2k^2}{2m\Lambda_{\bf k}},
 \label{20}
 \end{equation}
where $\hbar\omega_{\bf k}$ is the energy of excitation with
momentum $\hbar{\bf k}$, $\Lambda_{\bf k}$ is the structure factor
of superfluid helium. Within the framework of microscopic theory
\cite{15} using the approximation of pair interaction this
structure factor takes the form
 \begin{equation}
 \Lambda_{\bf k}= \left\{ 1+ 4
 \frac{\rho_LV({\bf k})}{\hbar^2k^2}
 \right\}^{-1/2}.
 \label{21}
 \end{equation}
This particularly means that if we know the correct excitation
spectrum of superfluid helium, say from experiment, then the
quantity $V({\bf k})$ can be unambiguously expressed through the
parameters of this spectrum. Namely
 \begin{equation}
 V({\bf k})=\frac{1}{\rho_L}\left(
 \frac{m^2\omega_{\bf k}^2}{k^2}-\frac{1}{4}\hbar^2k^2
 \right),
 \label{22}
 \end{equation}
where the analytic form of $\omega_{\bf k}$ depends on particular
range of wave vectors. In this manner the problem becomes
self-consistent.

 \par
The Fourier transforms of collective variables can now be written
in the second quantized form \cite{18}
 \begin{equation}
 n_{-\bf k}=-i\left(\frac{\rho_L\Lambda_{\bf k}}
 {m\Omega_L}\right)^{1/2}
 \left( {\hat a}_{\bf k}-
 {\hat a}_{-{\bf k}}^\dag\right)=
 -i\frac{\hbar k}{m}
 \left(
 \frac{\rho_L}{2\hbar\omega_{\bf k}\Omega_L}
 \right)^{1/2}
 \left( {\hat a}_{\bf k}-
 {\hat a}_{-{\bf k}}^\dag\right).
 \label{23}
 \end{equation}

 \begin{equation}
 {\bf v}_{\bf k}=-\frac{i}{2}\frac{\hbar{\bf k}}
 {\sqrt{m\rho_L\Omega_L\Lambda_{\bf k}}}
 \left( {\hat a}_{\bf k}+
 {\hat a}_{-{\bf k}}^\dag\right)=
 -i\left(
 \frac{\hbar\omega_{\bf k}}{2\rho_L\Omega_L}
 \right)^{1/2}
 \frac{\bf k}{k}
 \left( {\hat a}_{\bf k}+
 {\hat a}_{-{\bf k}}^\dag\right).
 \label{24}
 \end{equation}
where ${\hat a}_{\bf k}^\dag$ and ${\hat a}_{\bf k}$ are the
creation and annihilation operators of elementary excitation in
liquid helium with momentum $\hbar{\bf k}$ and energy $\omega_{\bf
k}$.

 \par
Now we must write down the $z$th component of lattice displacement
in the solid in terms of creation ${\hat b}_{{\bf q}_\alpha}^\dag$
and annihilation ${\hat b}_{{\bf q}_\alpha}$ operators for solid
phonons. Here ${\bf q}_\alpha$ is the wave vector of solid phonon
with $\alpha$th polarization. Taking into account the longitudinal
and transverse polarizations of bulk phonons and surface phonons
we can write
 \begin{eqnarray}
 u_z({\bf R})=\left(
 \frac{\hbar}{2\rho_S\Omega_S}
 \right)^{1/2}\Biggl\{
 \sum_{{\bf q}_l}\frac{B_l(\theta)}{\omega_{{\bf q}_l}^{1/2}}
 \left( {\hat b}_{{\bf q}_l}-{\hat b}_{-{\bf q}_l}^\dag\right)
 {\rm e}^{i{\bf q}_l{\bf R}} +
 \sum_{{\bf q}_t}\frac{B_t(\theta)}{\omega_{{\bf q}_t}^{1/2}}
 \left( {\hat b}_{{\bf q}_t}-{\hat b}_{-{\bf q}_t}^\dag\right)
 {\rm e}^{i{\bf q}_t{\bf R}} \nonumber \\ +
 \left(\frac{\Omega_S}{\Sigma} \right)^{1/2}
 \sum_{{\bf q}_s}\frac{B_s(q_s)}{\omega_{{\bf q}_s}^{1/2}}
 \left( {\hat b}_{{\bf q}_s}-{\hat b}_{-{\bf q}_s}^\dag\right)
 {\rm e}^{i{\bf q}_s{\bf r}_{\parallel}}
 \Biggr\},
 \label{25}
 \end{eqnarray}
where $\rho_S$ and $\Omega_S$ are the mass density and volume of
the solid, respectively, $\omega_{{\bf q}_\alpha}$ is the
frequency of $\alpha$th phonon mode, ${\bf q}_s$ is the
two-dimensional wave vector of surface phonon, ${\bf r}_\parallel
={\bf R}_\parallel$ is the parallel component of vector ${\bf r}$
(or ${\bf R}$) at $z=Z=0$. The angular amplitudes
$B_\alpha(\theta)$ can be determined as \cite{4}

 \begin{equation}
 B_l(\theta)=\frac{2s_l^2\cos{\theta_0}\cos{2\theta_t}}
 {s_t^2\sin{2\theta_t}\sin{2\theta_0}+s_l^2\cos^2{2\theta_t}},
 \label{26}
 \end{equation}

 \begin{equation}
 B_t(\theta)=\frac{2s_t^2\cos{\theta_0}\sin{2\theta_l}}
 {s_t^2\sin{2\theta_l}\sin{2\theta_0}+s_l^2\cos^2{2\theta_0}},
 \label{27}
 \end{equation}

 \begin{equation}
 B_s(q_s)=\frac{q_s^2-\kappa_t^2}{2q_s}g^{-1/2},
 \label{28}
 \end{equation}

 \begin{equation}
 g=\frac{q_s^2+\kappa_t^2}{2\kappa_t}+
 \frac{q_s^2+\kappa_t^2}{2\kappa_l}\frac{\kappa_t}{\kappa_l}
 - \frac{q_s^2+\kappa_t^2}{\kappa_l}
 \label{29}
 \end{equation}

 \begin{equation}
 \kappa_\alpha = \omega_{{\bf q}_\alpha}
 \sqrt{c^{-2}-s_\alpha^{-2}}, \quad (\alpha =l,t).
 \label{30}
 \end{equation}
Here $c$ is the velocity of first sound in superfluid helium,
$s_l$ and $s_t$ are the velocities of longitudinal and transverse
sounds in the solid, $\theta_0$ and $\theta_\alpha$ are the
incidence and reflection angles.

 \par
In fact, the formulae (\ref{14}), (\ref{18})--(\ref{21}), and
(\ref{25}) give us the sought transfer Hamiltonian in second
quantized form, which describes the transmission of elementary
excitations from the solid to superfluid helium.

 \par
Note, the resulting Hamiltonian contains the terms associated with
creation or annihilation of single excitation as well as pair of
excitations in helium, while it is only linear in creation and
annihilation operators of phonons in the solid. This asymmetry is
of course the shortcoming of our method. It can be avoided if we
take into account the third order anharmonisms in the solid.
However I am not concerned with this problem here because the main
goal of the present paper is to describe the transmission of
excitations from {\it heated solid} to the liquid helium at the
temperatures closed to zero, which is consistent with real
experimental situation \cite{14}. Under such conditions the
contribution of single solid phonon annihilation is expected to
dominate.

\section{Transfer Hamiltonians for particular processes}

In this Section I write down the Hamiltonians for particular
transfer processes: solid phonon $\Leftrightarrow$ single liquid
phonon (single phonon emission), solid phonon $\Leftrightarrow$
two liquid phonons (two phonons emission), solid phonon
$\Leftrightarrow$ single liquid roton (single roton emission).

\subsection{Single phonon emission}
To obtain the transfer Hamiltonian corresponding to the tunneling
of a single phonon from the solid to the liquid helium and {\it
vice versa} ({\it i.e.} to the processes ${\bf q}\leftrightarrow
{\bf k}$) we must retain in the expressions (\ref{18}) and
(\ref{19}) only liner terms with respect to helium collective
variable. In this case we get
 \begin{equation}
 \pi_{zz}^{(1{\rm ph})}=\sum_{\bf k}\left(
 \frac{\hbar^2k_z^2}{4m}+
 \frac{NV({\bf k})}{\Omega_L}
 \right)n_{-{\bf k}}{\rm e}^{i{\bf k}{\bf r}},
 \label{31}
 \end{equation}
where I have used $n_0=N/\Omega_L$, $N$ is the total number of
helium atoms. In the case of helium phonons $\omega_{\bf k} =
c{\bf k}$. Thus the relation (\ref{22}) yields
 \begin{equation}
 V({\bf k})=\frac{\Omega_L}{N}\left(
 mc^2-\frac{\hbar^2k^2}{4m}\right)\approx
 \frac{\Omega_L}{N}mc^2,
 \label{32}
 \end{equation}
where I have neglected the term quadratic in wave vector. The same
approximation can be done in the first term of eq. (\ref{31}). In
fact, without these approximations one can see that after
substitution of eq. (\ref{32}) into eq. (\ref{31}) both small
terms give even smaller one $\sim(k^2-k_z^2)$, because $k\sim k_z$
in the process of interest. So, using the relations (\ref{32}) and
(\ref{31}) we come to
 \begin{equation}
 \pi_{zz}^{(1{\rm ph})}=mc^2\sum_{\bf k}
 n_{-{\bf k}}{\rm e}^{i{\bf k}{\bf r}}.
 \label{33}
 \end{equation}
Substituting eqs. (\ref{25}) and (\ref{33}) into eq. (\ref{14})
and integrating over the interface area $\Sigma$ we obtain the
transfer Hamiltonian for the single phonon channel in the
following form
 \begin{eqnarray}
 \hat{\cal H}_{\rm{int}}^{(1{\rm ph})}=
 i\Sigma c\hbar \left(
 \frac{\rho_L}{\rho_S\Omega_S\Omega_L}
 \right)^{1/2}
 \Biggl\{
 \sum_{{\bf k},{\bf q}_l}B_l(\theta)
 \left( \frac{\omega_{\bf k}}{\omega_{{\bf q}_l}}\right)^{1/2}
 \left(
 \hat{a}_{\bf k}\hat{b}_{{\bf q}_l}^\dag -
 \hat{a}_{\bf k}^\dag\hat{b}_{{\bf q}_l}
 \right)\delta_{{\bf k}_\parallel ,{\bf q}_{l\parallel}}
 \nonumber \\+
 \sum_{{\bf k},{\bf q}_t}B_t(\theta)
 \left( \frac{\omega_{\bf k}}{\omega_{{\bf q}_t}}\right)^{1/2}
 \left(
 \hat{a}_{\bf k}\hat{b}_{{\bf q}_t}^\dag -
 \hat{a}_{\bf k}^\dag\hat{b}_{{\bf q}_t}
 \right)\delta_{{\bf k}_\parallel ,{\bf q}_{t\parallel}}
 \nonumber \\ +
 \left( \frac{\Omega_S}{\Sigma}\right)^{1/2}
 \sum_{{\bf k},{\bf q}_s}B_s(q_s)
 \left( \frac{\omega_{\bf k}}{\omega_{{\bf q}_s}}\right)^{1/2}
 \left(
 \hat{a}_{\bf k}\hat{b}_{{\bf q}_s}^\dag -
 \hat{a}_{\bf k}^\dag\hat{b}_{{\bf q}_s}
 \right)\delta_{{\bf k}_\parallel ,{\bf q}_s}
 \Biggr\}.
 \label{34}
 \end{eqnarray}
This formula will be used in Section IV for deriving the
Khalatnikov formula for heat flow via the single phonon channel.

\subsection{Two phonons emission}

The obtained formulae (\ref{14}), (\ref{18}), and (\ref{19}) allow
us to study also the processes of emission of two helium phonons
from single phonon in the solid. For this purpose let us retain in
eqs. (\ref{18}) and (\ref{19}) only the terms quadratic in
collective variables of liquid helium. This leads to the following
expression for stress tensor
 \begin{equation}
 \pi_{zz}^{(2{\rm ph})}({\bf r})=
 \sum_{{\bf k},{\bf k}'}\left\{
 \rho_Lv_{{\bf k}z}v_{{\bf k}'z}+
 \frac{V({\bf k})k_z}{k_z+k'_z}
 n_{-{\bf k}}n_{-{\bf k}'}
 \right\}{\rm e}^{i({\bf k}+{\bf k}'){\bf r}},
 \label{35}
 \end{equation}
which gives the transfer Hamiltonian for the two phonons processes
under consideration in the form
 \begin{eqnarray}
 \hat{\cal H}_{\rm{int}}^{(2{\rm ph})}=
 \frac{2\hbar\Sigma}{\Omega_L} \left(
 \frac{\hbar}{2\rho_S\Omega_S}
 \right)^{1/2}
 \Biggl\{
 \sum_{{\bf k}, {\bf k}', {\bf q}_l}B_l(\theta)
 \left( \frac{\omega_{\bf k}\omega_{{\bf k}'}}
 {\omega_{{\bf q}_l}}\right)^{1/2}
 \Xi ({\bf k},{\bf k}',{\bf q}_l)
 \nonumber \\+
 \sum_{{\bf k}, {\bf k}', {\bf q}_t}B_t(\theta)
 \left( \frac{\omega_{\bf k}\omega_{{\bf k}'}}
 {\omega_{{\bf q}_t}}\right)^{1/2}
 \Xi ({\bf k},{\bf k}',{\bf q}_t)
 \nonumber \\ +
 \left( \frac{\Omega_S}{\Sigma}\right)^{1/2}
 \sum_{{\bf k}, {\bf k}', {\bf q}_s}B_s(q_s)
 \left( \frac{\omega_{\bf k}\omega_{{\bf k}'}}
 {\omega_{{\bf q}_s}}\right)^{1/2}
 \Xi ({\bf k},{\bf k}',{\bf q}_s)
 \Biggr\}.
 \label{36}
 \end{eqnarray}
Here I have introduced the functions
 \begin{equation}
 \Xi ({\bf k},{\bf k}',{\bf q}_\alpha) =
 {\cal J}_1 ({\bf q}_\alpha ,{\bf k}\vert{\bf k}')
 \delta_{{\bf q}_{\alpha\parallel}+{\bf k}_\parallel,
 {\bf k}'_\parallel} -
 {\cal J}_2 ({\bf q}_\alpha\vert{\bf k},{\bf k}')
 \delta_{{\bf q}_{\alpha\parallel},{\bf k}_\parallel+
 {\bf k}'_\parallel},
 \label{37}
 \end{equation}
where
 \begin{equation}
 {\cal J}_1 ({\bf q}_\alpha ,{\bf k}\vert{\bf k}')=
 \left( \frac{1}{2}+ \frac{k_zk'_z}{kk'} \right)
 \left( \hat{b}_{{\bf q}_\alpha}\hat{a}_{\bf k}
 \hat{a}_{{\bf k}'}^\dag +
 \hat{b}_{{\bf q}_\alpha}^\dag\hat{a}_{\bf k}^\dag
 \hat{a}_{{\bf k}'}\right),
 \label{38}
 \end{equation}

 \begin{equation}
 {\cal J}_2 ({\bf q}_\alpha \vert{\bf k},{\bf k}')=
 \frac{1}{2}\left( \frac{k_zk'_z}{kk'} +
 \frac{k_z}{k_z+k'_z}\right)
 \left( \hat{b}_{{\bf q}_\alpha}\hat{a}_{\bf k}^\dag
 \hat{a}_{{\bf k}'}^\dag +
 \hat{b}_{{\bf q}_\alpha}^\dag\hat{a}_{\bf k}
 \hat{a}_{{\bf k}'}\right).
 \label{39}
 \end{equation}
Formulae (\ref{36}) -- (\ref{39}) provide a possibility to study
the energy transfer through the solid -- liquid helium interface
without conservation of number of quasiparticles. In this case
there is no strong restriction on the angles for created phonons.
Therefore such processes can lead to the isotropic background,
which has been observed in experiments on phonon emission from
cleaved crystals \cite{9,10}. This problem will be considered in
detail in another paper.

\subsection{Single roton emission}

To obtain the tunnel Hamiltonian, which corresponds to creation of
a single roton in He II by a solid phonon, we can again use the
relation (\ref{31}) with account of eq. (\ref{22}). In this case
we however cannot neglect the terms quadratic in wave vectors.
Note, the frequency $\omega_{\bf k}$ in (\ref{31}) represents now
the roton spectrum, whose analytic form depends on the chosen
concrete model. Here I do not distinguish the R$^{(-)}$ and
R$^{(+)}$ rotons. This problem will be considered in detail in
Section V.

 \par
Substituting eq. (\ref{22}) into eq. (\ref{31}) we get
 \begin{equation}
 \pi_{zz}^{(1{\rm R})}({\bf r})=\sum_{\bf k}
 \left\{
 \frac{m\omega_{\bf k}^2}{k^2}+\frac{\hbar^2(k_z^2-k^2)}{4m}
 \right\}n_{-{\bf k}}{\rm e}^{i{\bf k}{\bf r}}.
 \label{40}
 \end{equation}
Finally the corresponding transfer Hamiltonian takes the form
 \begin{eqnarray}
 \hat{\cal H}_{\rm{int}}^{(1{\rm R})}=
 \frac{i\Sigma\hbar}{2m} \left(
 \frac{\rho_L}{\rho_S\Omega_S\Omega_L}
 \right)^{1/2}
 \Biggl\{
 \sum_{{\bf k},{\bf q}_l}
 \frac{B_l(\theta){\cal F}({\bf k})k}
 {({\omega_{\bf k}}{\omega_{{\bf q}_l}})^{1/2}}
 \left(
 \hat{a}_{\bf k}\hat{b}_{{\bf q}_l}^\dag -
 \hat{a}_{\bf k}^\dag\hat{b}_{{\bf q}_l}
 \right)\delta_{{\bf k}_\parallel ,{\bf q}_{l\parallel}}
 \nonumber \\+
 \sum_{{\bf k},{\bf q}_t}
 \frac{B_t(\theta){\cal F}({\bf k})k}
 {({\omega_{\bf k}}{\omega_{{\bf q}_t}})^{1/2}}
 \left(
 \hat{a}_{\bf k}\hat{b}_{{\bf q}_t}^\dag -
 \hat{a}_{\bf k}^\dag\hat{b}_{{\bf q}_t}
 \right)\delta_{{\bf k}_\parallel ,{\bf q}_{t\parallel}}
 \nonumber \\ +
 \left( \frac{\Omega_S}{\Sigma}\right)^{1/2}
 \sum_{{\bf k},{\bf q}_s}\frac{B_s(q_s){\cal F}({\bf k})k}
 {({\omega_{\bf k}}{\omega_{{\bf q}_s}})^{1/2}}
 \left(
 \hat{a}_{\bf k}\hat{b}_{{\bf q}_s}^\dag -
 \hat{a}_{\bf k}^\dag\hat{b}_{{\bf q}_s}
 \right)\delta_{{\bf k}_\parallel ,{\bf q}_s}
 \Biggr\},
 \label{41}
 \end{eqnarray}
where
 \begin{equation}
 {\cal F}({\bf k})=\left\{
 \frac{m\omega_{\bf k}^2}{k^2}+\frac{\hbar^2(k_z^2-k^2)}{4m}
 \right\}.
 \label{42}
 \end{equation}
In Section V I will use the Hamiltonian (\ref{41}) for derivation
of energy flux through the interface via the single roton channel.

 \par
As the curtain fell of this section it should be noted that using
the general formulae (\ref{18}) and (\ref{19}) we could study also
various inelastic processes with rotons, {\it e.g.} solid phonon
$\Leftrightarrow$ two rotons, solid phonon $\Leftrightarrow$ roton
$+$ phonon in helium, and scattering of roton on the surface with
creation of a phonon in the solid.

\section{Single phonon emission. Derivation of the Khalatnikov
formula}

Any new theory claiming to describe the emission of elementary
excitations from the solid to the superfluid helium should first
be tested on the basic result obtained by Khalatnikov for a heat
flow via the single phonon channel \cite{4}. Really, this latter
result was derived within appropriate framework, and is in
excellent agreement with the experimental observations \cite{9,7}.

 \par
To calculate the energy flux through the solid--liquid interface
via single phonon channel we must first calculate the probability
density function defined by the usual relation
 \begin{equation}
 W^{(1{\rm ph})}=\frac{2\pi}{\Sigma\hbar^2}\left\vert
 {\cal M}_{{\bf q}\leftrightarrow{\bf k}}^{(1{\rm ph})}
 \right\vert^2\delta(\omega_{\bf q}-\omega_{\bf k}),
 \label{43}
 \end{equation}
where ${\cal M}_{{\bf q}\leftrightarrow{\bf k}}^{(1{\rm ph})} =
\langle{\bf q}|\hat{\cal H}_{\rm{int}}^{(1{\rm ph})}| {\bf k}
\rangle$ is the corresponding matrix element of the transfer
Hamiltonian given by eq. (\ref{34}). Substituting (\ref{34}) into
(\ref{43}) we easily get
 \begin{equation}
 W^{(1{\rm ph})}=(2\pi)^3 \left\{
 {\cal A}_l^{(1{\rm ph})}
 \delta({\bf k}_\parallel-{\bf q}_{l\parallel})+
 {\cal A}_t^{(1{\rm ph})}
 \delta({\bf k}_\parallel-{\bf q}_{t\parallel})+
 {\cal A}_s^{(1{\rm ph})}
 \delta({\bf k}_\parallel-{\bf q}_{s})
 \right\}\delta(\omega_{\bf q}-\omega_{\bf k}),
 \label{44}
 \end{equation}
where the coefficients ${\cal A}_\alpha$ are defined by the
relations
 \[
 {\cal A}_l^{(1{\rm ph})}=
 \frac{\omega_{\bf k}}{\omega_{{\bf q}_l}}
 \frac{\rho_Lc^2}{\rho_S\Omega_L\Omega_S}B_l^2(\theta), \quad
 {\cal A}_t^{(1{\rm ph})}
 =\frac{\omega_{\bf k}}{\omega_{{\bf q}_t}}
 \frac{\rho_Lc^2}{\rho_S\Omega_L\Omega_S}B_t^2(\theta), \quad
 {\cal A}_s^{(1{\rm ph})}
 =\frac{\omega_{\bf k}}{\omega_{{\bf q}_s}}
 \frac{\rho_Lc^2}{\rho_S\Omega_L\Sigma}B_s^2(q_s).
 \]
To obtain formula (\ref{44}) I have done the transition from
discrete to continuous wave vectors:
 \[
 \delta_{{\bf k}_\parallel ,{\bf q}_{\alpha\parallel}}
 \longrightarrow \frac{(2\pi)^2}{\Sigma}
 \delta ({\bf k}_\parallel -{\bf q}_{\alpha\parallel}).
 \]
The probability of creation of phonon with group velocity
 \begin{equation}
 {\bf c}\approx \frac{\omega_{\bf k}}{k}\frac{\bf k}{k}
 \label{45}
 \end{equation}
near the surface element $d{\bf \Sigma}\parallel {\bf e}_z$ (${\bf
e}_z$ is the unit vector along $z$th direction) is given by
 \begin{equation}
 W^{(1{\rm ph})}\hbar\omega_{\bf k}n(T_S)
 \frac{\left({\bf c}\cdot d{\bf \Sigma} \right)}{c}
 \approx W^{(1{\rm ph})}\hbar\omega_{\bf k}n(T_S) d\Sigma,
 \label{46}
 \end{equation}
where $n(T_S)=\left[\exp(\hbar\omega_{{\bf
q}_\alpha}/k_bT_S)-1\right]^{-1}$ is the equilibrium distribution
function of phonons in the solid, $T_S$ is the temperature of the
solid. Thus, the heat flux from solid to liquid helium per unit
area can be written as follows
 \begin{eqnarray}
 Q_{S\rightarrow L}^{(1{\rm ph})}= (2\pi)^3 \Biggl\{
 \int{\cal A}_l\hbar\omega_{\bf k}n(T_S)
 \delta ({\bf k}_\parallel -{\bf q}_{l\parallel})
 \delta(\omega_{{\bf q}_l}-\omega_{\bf k})
 \frac{\Omega_Ld^3k}{(2\pi)^3}\frac{\Omega_Sd^3q_l}{(2\pi)^3}
 \nonumber \\ +
 \int{\cal A}_t\hbar\omega_{\bf k}n(T_S)
 \delta ({\bf k}_\parallel -{\bf q}_{t\parallel})
 \delta(\omega_{{\bf q}_t}-\omega_{\bf k})
 \frac{\Omega_Ld^3k}{(2\pi)^3}\frac{\Omega_Sd^3q_t}{(2\pi)^3}
 \nonumber \\ +
 \int{\cal A}_s\hbar\omega_{\bf k}n(T_S)
 \delta ({\bf k}_\parallel -{\bf q}_s)
 \delta(\omega_{{\bf q}_s}-\omega_{\bf k})
 \frac{\Omega_Ld^3k}{(2\pi)^3}\frac{\Sigma d^2q_s}{(2\pi)^2}
 \Biggr\}.
 \label{47}
 \end{eqnarray}
Delta functions of the type $\delta({\bf k}_\parallel-{\bf
q}_\alpha)$ take away integrations over ${\bf k}_\parallel$. Delta
functions $\delta(\omega_{{\bf q}_\alpha}-\omega_{\bf k})$ can be
excluded by the integration over $k_z$ (note, $k_z\approx k$).
After these straightforward calculations the integrals over the
energies and over angles can be separated, so that the final
result takes the form
 \begin{equation}
 Q_{S\rightarrow L}^{(1{\rm ph})} =
 \frac{4\pi\hbar}{(2\pi)^3}\frac{\rho_Lc}{\rho_S}
 \frac{F(\eta)}{s_t^3}\int\limits_0^\infty
 \frac{\omega^3d\omega}
 {\exp\left(\frac{\hbar\omega}{k_bT_S}\right)-1}=
 \frac{\pi^2}{30}\frac{\rho_Lc}{\hbar^3\rho_Ss_t^3}
 (k_bT_S)^4 F(\eta),
 \label{48}
 \end{equation}
where $F(\eta)$ is the function of $\eta=s_l/s_t$ written down in
Ref. 4. The formula (\ref{48}) exactly coincides with the well
known Khalatnikov result \cite{4} derived initially in the
framework of acoustic mismatch theory. Note, the total heat flux
between the solid and liquid helium is given by the difference
 \begin{equation}
 J^{(1{\rm ph})}= Q_{S\rightarrow L}^{(1{\rm ph})}-
 Q_{L\rightarrow S}^{(1{\rm ph})},
 \label{49}
 \end{equation}
where $Q_{L\rightarrow S}^{(1{\rm ph})}$ can be found in the
similar way. Finally it is given by the same formula as (\ref{48})
up to the replacement $T_S\longrightarrow T_L$, where $T_L$ is the
temperature of liquid helium.

\section{Rotons emission}

The novel techniques developed during last decade provide a
possibility to produce and unambiguously detect by quantum
evaporation processes both the R$^{(-)}$ rotons ($k<k_0$) and the
R$^{(+)}$ rotons ($k>k_0$). However, it was realized that simple
flat metal films produce easily measurable evaporation signals due
to R$^{(+)}$ rotons \cite{20}, but no signal from R$^{(-)}$ rotons
has ever been identified from such sources. To produce R$^{(-)}$
rotons two types of sources have been developed \cite{21,22} ,
which are based on trapping of the emitted R$^{(+)}$ rotons to
enable R$^{(-)}$ rotons to be created by interactions in
nonequilibrium gas of excitations. Furthermore, Tucker and Wyatt
reported \cite{22} about creation of fast enough pulsed source of
R$^{(-)}$ rotons suitable for time of flight measurements.

 \par
Let us at first demonstrate the impossibility of direct emission
of R$^{(-)}$ rotons from a solid, and, secondly, to describe the
process of R$^{(+)}$ rotons emission.

\subsection{R$^{(-)}$ rotons}

It is easy to see that R$^{(-)}$ rotons and R$^{(+)}$ rotons are
not principally distinguished neither in the stress tensor
(\ref{40}) nor in the corresponding transfer Hamiltonian
(\ref{41}). Really the probability of R$^{(-)}$ roton emission
seems to be nonzero if the energy and parallel component of
momentum are conserved. This is just the case in the emission of
real particles, say quantum evaporation \cite{23}. But
quasiparticles in media are the {\it collective} excitations,
which originated from local spatial perturbations. In this case
the conservation laws at the surface should be obtained directly
from the boundary conditions. In fact, the boundary conditions
(\ref{6}) and (\ref{7}) just led us to the interaction Hamiltonian
(\ref{8}) we have used. But the reversed conditions (which are
clearly absurd) will necessarily give the same expression
(\ref{8})! This means that for accurate use of this interaction
Hamiltonian we must always keep in mind at least one of the
boundary conditions, say (\ref{6}) (the second condition (\ref{7})
is automatically satisfied as long as we use the transfer
Hamiltonian (\ref{8})). In original SBT theory \cite{13} there
were no boundary conditions at all. This is the reason why the
problem of R$^{(-)}$ rotons creation even did not appear in Ref.
13.

 \par
Let me now analyze the first boundary condition (\ref{6}). In view
of the expression (\ref{24}) the normal component of displacement
$\delta r_z(x,y,0)$ at the surface can be written as follows
\cite{24}
 \begin{equation}
 \delta r_z(x,y,0)=\left(\frac{\hbar}{2\rho_L\Omega_L}\right)^{1/2}
 \sum_{\bf k}\omega_{\bf k}^{-1/2}\frac{k_z}{k}
 \left( {\hat a}_{\bf k}-
 {\hat a}_{-{\bf k}}^\dag\right)
 {\rm e}^{i{\bf k_\parallel}{\bf r_\parallel}},
 \label{50}
 \end{equation}
while retaining for simplicity one polarization, say longitudinal,
eq. (\ref{25}) at the interface can be rewritten in the analogous
form:
 \begin{equation}
 u_z(x,y,0)=\left(
 \frac{\hbar}{2\rho_S\Omega_S}
 \right)^{1/2}
 \sum_{\bf q}\omega_{\bf q}^{-1/2}
 \frac{q_z}{q}
 \left( {\hat b}_{\bf q}-{\hat b}_{-{\bf q}}^\dag\right)
 {\rm e}^{i{\bf q}_\parallel{\bf R}_\parallel}.
 \label{51}
 \end{equation}
The right hand sides of eqs. (\ref{50}) and (\ref{51}) can be
equal only if all their matrix elements in arbitrary basis are
equal. Just this requirement leads to the conservation of the
component of momentum parallel to the interface. But this can only
occur if the coefficients at exponents in eqs. (\ref{50}) and
(\ref{51}) have the same sign!

 \par
The detectable quasiparticle propagating into He II must have the
positive $z$th component of group velocity. As is known momentum
of R$^{(-)}$ rotons is oppositely directed to their group
velocity. This results in impossibility to satisfy the boundary
condition (\ref{6}) in view of eqs. (\ref{50}) and (\ref{51}).

\subsection{R$^{(+)}$ rotons emission}

Let us now concentrate our attention in the processes of R$^{(+)}$
rotons emission. In analogy with the previous section we can
calculate the probability density function for single roton
creation as
 \[
 W^{(1{\rm R})}=\frac{2\pi}{\Sigma\hbar^2}\left\vert
 {\cal M}_{{\bf q}\leftrightarrow{\bf k}}^{(1{\rm R})}
 \right\vert^2\delta(\omega_{\bf q}-\omega_{\bf k})
 \]

 \begin{equation}
 = (2\pi)^3 \left\{
 {\cal A}_l^{(1{\rm R})}
 \delta({\bf k}_\parallel-{\bf q}_{l\parallel})+
 {\cal A}_t^{(1{\rm R})}
 \delta({\bf k}_\parallel-{\bf q}_{t\parallel})+
 {\cal A}_s^{(1{\rm R})}
 \delta({\bf k}_\parallel-{\bf q}_{s})
 \right\}\delta(\omega_{\bf q}-\omega_{\bf k}),
 \label{52}
 \end{equation}
where ${\bf k}$ denotes now the roton wave vector, $\omega_{\bf
k}$ is its energy. The coefficients ${\cal A}_l^{(1{\rm R})}$ are
determined by the relations
 \[
 {\cal A}_l^{(1{\rm R})}=
 \frac{\rho_L}{\rho_S\Omega_L\Omega_S}
 \frac{k^2}{4m^2} \frac{B_l^2(\theta)F^2({\bf k})}
 {\omega_{\bf k}\omega_{{\bf q}_l}}, \quad
 {\cal A}_t^{(1{\rm R})}=
 \frac{\rho_L}{\rho_S\Omega_L\Omega_S}
 \frac{k^2}{4m^2} \frac{B_t^2(\theta)F^2({\bf k})}
 {\omega_{\bf k}\omega_{{\bf q}_t}},
 \]

 \[
 {\cal A}_s^{(1{\rm R})}=
 \frac{\rho_L}{\rho_S\Omega_L\Sigma}
 \frac{k^2}{4m^2} \frac{B_s^2(q_s)F^2({\bf k})}
 {\omega_{\bf k}\omega_{{\bf q}_s}}.
 \]
The probability of creation of roton with energy $\hbar\omega_{\bf
k }$ and group velocity
 \[
 {\bf v}_r=\frac{\partial\omega_{\bf k}}{\partial{\bf k}}
 \]
near the surface element $d{\bf \Sigma}\parallel {\bf e}_z$ is
equal to
 \begin{equation}
 W^{(1{\rm R})}\hbar\omega_{\bf k}n(T_S)
 \frac{\left({\bf v}_r\cdot d{\bf \Sigma} \right)}{v_r}
 = W^{(1{\rm R})}\hbar\omega_{\bf k}n(T_S) \cos\theta d\Sigma,
 \label{53}
 \end{equation}
where $\cos\theta ={\bf v}_r\cdot{\bf e}_z$. Therefore, the energy
flux per unit area associated with single roton channel takes the
form
 \begin{eqnarray}
 Q_{S\rightarrow L}^{(1{\rm R})}= (2\pi)^3 \Biggl\{
 \int{\cal A}_l^{(1{\rm R})}\hbar\omega_{\bf k}\cos\theta n(T_S)
 \delta ({\bf k}_\parallel -{\bf q}_{l\parallel})
 \delta(\omega_{{\bf q}_l}-\omega_{\bf k})
 \frac{\Omega_Ld^3k}{(2\pi)^3}\frac{\Omega_Sd^3q_l}{(2\pi)^3}
 \nonumber \\ +
 \int{\cal A}_t^{(1{\rm R})}\hbar\omega_{\bf k}\cos\theta n(T_S)
 \delta ({\bf k}_\parallel -{\bf q}_{t\parallel})
 \delta(\omega_{{\bf q}_t}-\omega_{\bf k})
 \frac{\Omega_Ld^3k}{(2\pi)^3}\frac{\Omega_Sd^3q_t}{(2\pi)^3}
 \nonumber \\ +
 \int{\cal A}_s^{(1{\rm R})}\hbar\omega_{\bf k}\cos\theta n(T_S)
 \delta ({\bf k}_\parallel -{\bf q}_s)
 \delta(\omega_{{\bf q}_s}-\omega_{\bf k})
 \frac{\Omega_Ld^3k}{(2\pi)^3}\frac{\Sigma d^2q_s}{(2\pi)^2}
 \Biggr\}.
 \label{54}
 \end{eqnarray}
In principle the partial integrations in eq. (\ref{54}) can give
the spectral and angular distributions of emitted rotons. Here I
calculate the total energy yield from single roton channel. The
account of all phonon polarizations in the solid can only give the
multiplier similar to $F(\eta)$ appeared in eq. (\ref{48}). This
multiplier is of the order of unity for majority of solids. The
account of all polarizations can have an essential influence only
on the angular distribution. So, to calculate the total energy
flux let us restrict ourselves to one polarization, say
longitudinal one. Further I drop all the polarization subscripts.
Thus, the energy flux of interest looks
 \begin{eqnarray}
 Q_{S\rightarrow L}^{(1{\rm R})} =
 \frac{\hbar\rho_L}{4(2\pi)^3m^2\rho_S}\int
 \frac{k^2\omega_{\bf q}\omega_{\bf k}^{-2}\cos\theta}
 {\exp\left(\frac{\hbar\omega_{\bf q}}{k_bT_S}\right)-1}
 \left\{
 \frac{m\omega_{\bf k}^2}{k^2}+\frac{\hbar^2
 \left( k_z^2-k^2\right)}{4m}
 \right\}^2
 \nonumber \\ \times
 \delta({\bf q}_\parallel-{\bf k}_\parallel)
 \delta(\omega_{\bf q}-\omega_{\bf k})n(T_S)d^3kd^3q.
 \label{55}
 \end{eqnarray}
After the straightforward, but quite cumbersome calculations eq.
(\ref{55}) can be reduced to the single integral. Let me write
down here the spectral distribution for rotons emission:
 \begin{equation}
 \frac{dQ_{S\rightarrow L}^{(1{\rm R})}}{dk} =
 \frac{\rho_L\hbar^5}{120(2\pi)^2m^4s^7\rho_S}
 \frac{k^2\omega_{\bf k}^5}
 {\exp\left(\frac{\hbar\omega_{\bf k}}{k_bT_S}\right)-1}
 \left\{
 1-10\frac{m^2s^2}{\hbar^2k^2}+30\frac{m^4s^4}{\hbar^4k^4}
 \right\},
 \label{56}
 \end{equation}
To obtain the total energy flux via the single roton channel
$Q_{S\rightarrow L}^{(1{\rm R})}$, the relation (\ref{56}) should
be integrated over $k$ in the range $\{k_0,k_{\rm max}\}$, where
$k_{\rm max}$ is the maximum roton wave vector. Note again that
the formula (\ref{56}) gives the energy flux in the case $T_L=0$.
In general, $Q_{L\rightarrow S}^{(1{\rm R})}$ can be calculated in
the same way.

 \par
To make a numerical analysis of the obtained results we must
choose the particular analytic expression for roton spectrum in He
II. Precise numerical calculations require of course the analytic
expression most closed to a reality. But to draw the general
picture I will use here the classical Landau expression \cite{1}
 \begin{equation}
 \hbar\omega_{\bf k}=\Delta +\frac{\hbar^2(k-k_0)^2}{2\mu}
 \label{57}
 \end{equation}
where $\Delta\approx 8.6$ K, $k_0\approx 1.9\cdot 10^8$ cm$^{-1}$,
$\mu\approx 0.13m_{\rm He^4}\approx 0.868\cdot 10^{-24}$ g.

 \par
Figure \ref{f2} presents the spectral distribution of roton
emission calculated from (\ref{56}) for gold. Figure clearly shows
that the maximum at $k>k_0$ appears when temperature of the solid
exceed $2$ K. Integrating eq. (\ref{56}) over R$^{(+)}$ rotons
momenta we get the total energy flux. The ratio of this flux
associated with the single roton processes and that corresponding
to the single phonon processes given by eq. (\ref{48}) is shown in
Figure \ref{f2} as a function of temperature of the solid (again
gold). Previously SBT \cite{13} had also estimated the
contribution of single roton emission processes into the total
energy transmission through the solid -- liquid helium interface.
They, however, considered the case of small differences between
$T_S$ and $T_L$. Corresponding formulae can be easily obtained
from eqs. (\ref{56}) and (\ref{48}) by differentiation with
respect to a temperature and subsequent integration over roton and
phonon momenta, respectively. The corresponding ratio is shown in
Figure \ref{f3} together with SBT result. The estimation by SBT
\cite{13} gives the ratio approximately three times larger for
$T_S=2$ K. This difference is caused by the approximations done in
Ref. 13. Particularly the authors had taken $k_z\approx k$, which
is far from reality for rotons. Additionally they included
R$^{(-)}$ rotons in their calculations, which is incorrect as I
show in the previous subsection. But one can see from Fig.
\ref{f3} that their rough estimation nevertheless show right
qualitative behaviour of the energy flux due to roton emission.

\section{Conclusions}

I studied the emission of elementary excitations into liquid
helium II from a heated solid. For this process I developed the
new microscopic quantum theory generalizing the previous theory
\cite{13}. To describe the coupling between the solid and He II
the correct expression for the interface Hamiltonian (\ref{8}) has
been derived. This expression follows from the general principles
of quantum mechanics with account of boundary conditions at
solid--liquid interface (\ref{6}) and (\ref{7}). Following the
ideas of Ref. 13 and using the microscopic description of
superfluid helium \cite{15} the transfer Hamiltonian (\ref{8}) has
been expressed in terms of creation and annihilation operators of
elementary excitations in the solid and helium II. The transfer
Hamiltonians are explicitly written down for i) single phonon
emission (\ref{34}), ii) two phonons emission (\ref{36}), and iii)
single roton emission (\ref{41}). The heat flux trough the
interface associated with the single phonon channel is calculated
by the use of the derived microscopic expression (\ref{34}). The
obtained formula (\ref{48}) exactly coincides with the
corresponding result obtained by Khalatnikov \cite{4} using the
acoustic-mismatch approach. The heat flux associated with the
single R$^{(+)}$ roton channel (see formulas (\ref{55}) and
(\ref{56})) has been calculated as well. It is shown that the
boundary conditions (\ref{6}) and (\ref{7}) forbid the emission of
R$^{(-)}$ rotons by a solid. This latter result is in accordance
with experimental observations.

 \acknowledgments
I would like to acknowledge the numerous stimulating discussions
on the subject with Prof. I. N. Adamenko and Dr. K. {\'E}.
Nemchenko. I wish to express also my deep gratitude to Prof.
S.~Yamasaki and Prof. M.~Hirooka who acquainted me with the papers
cited in Ref. 15.

\begin{figure}
\caption{Geometry of the problem} \label{f1}
\end{figure}

\begin{figure}
\caption{Spectral distribution of roton emission into He II from
golden heater calculated by eq. (\ref{56})} \label{f2}
\end{figure}

\begin{figure}
\caption{Ratio of energy flux associated with the single roton
channel and that corresponding to the single phonon (Khalatnikov)
channel in case $T_L=0$} \label{f3}
\end{figure}

\begin{figure}
\caption{Ratio of energy flux associated with the single roton
channel and that corresponding to the single phonon (Khalatnikov)
channel in case $T_L\sim T_S$. Corresponding result of SBT
\cite{13} is presented by the broken curve} \label{f4}
\end{figure}


\begin{references}
\bibitem{18} The expressions (\ref{23}) and (\ref{24}) are
             different in form from those in Refs. 13 and 15,
             because of using another notations. Such a method
             of presentation is closer to the ordinary
             quantization of liquid helium hydrodynamic
             Hamiltonian (see the first paper in Ref. 8).
\end{references}
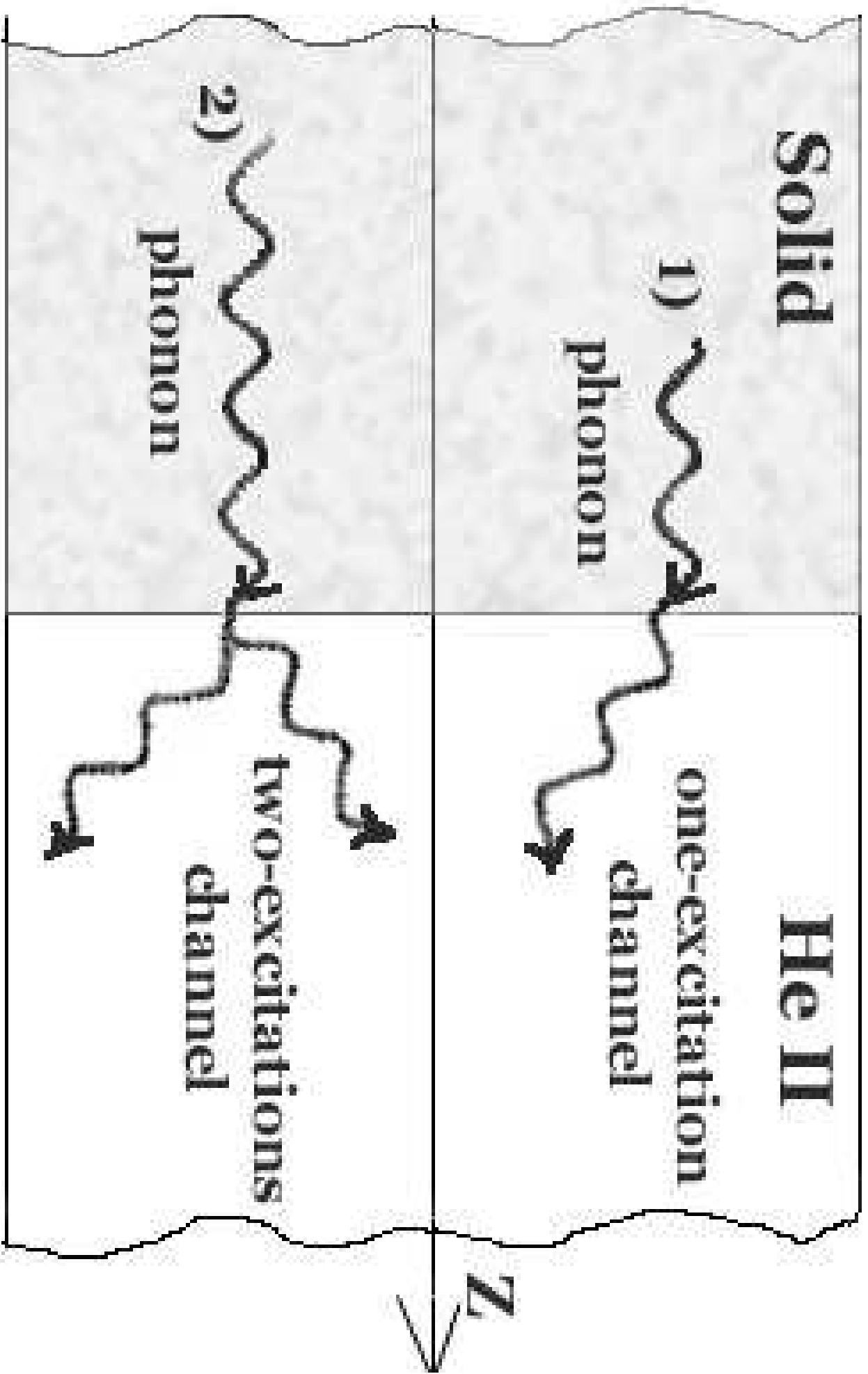
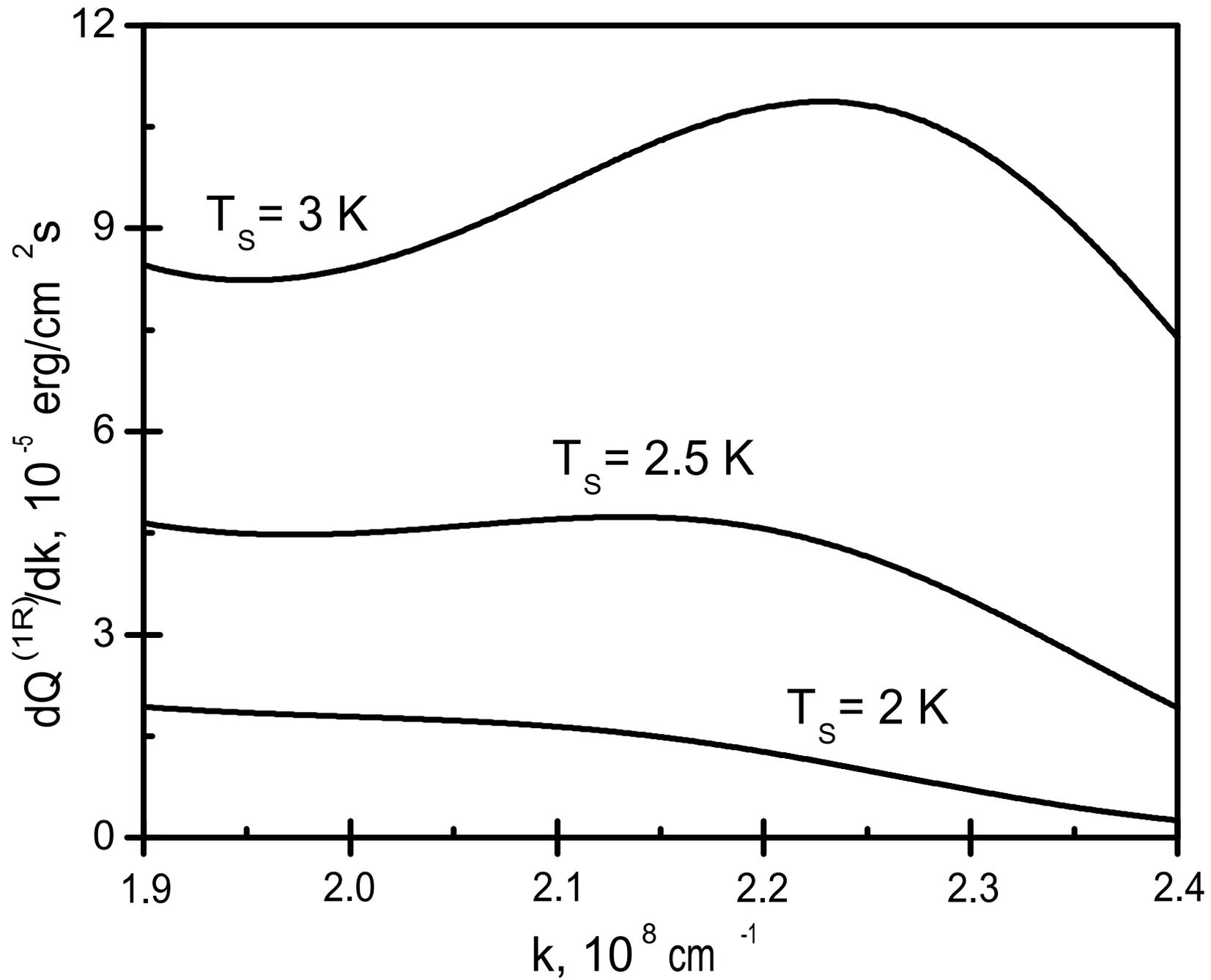
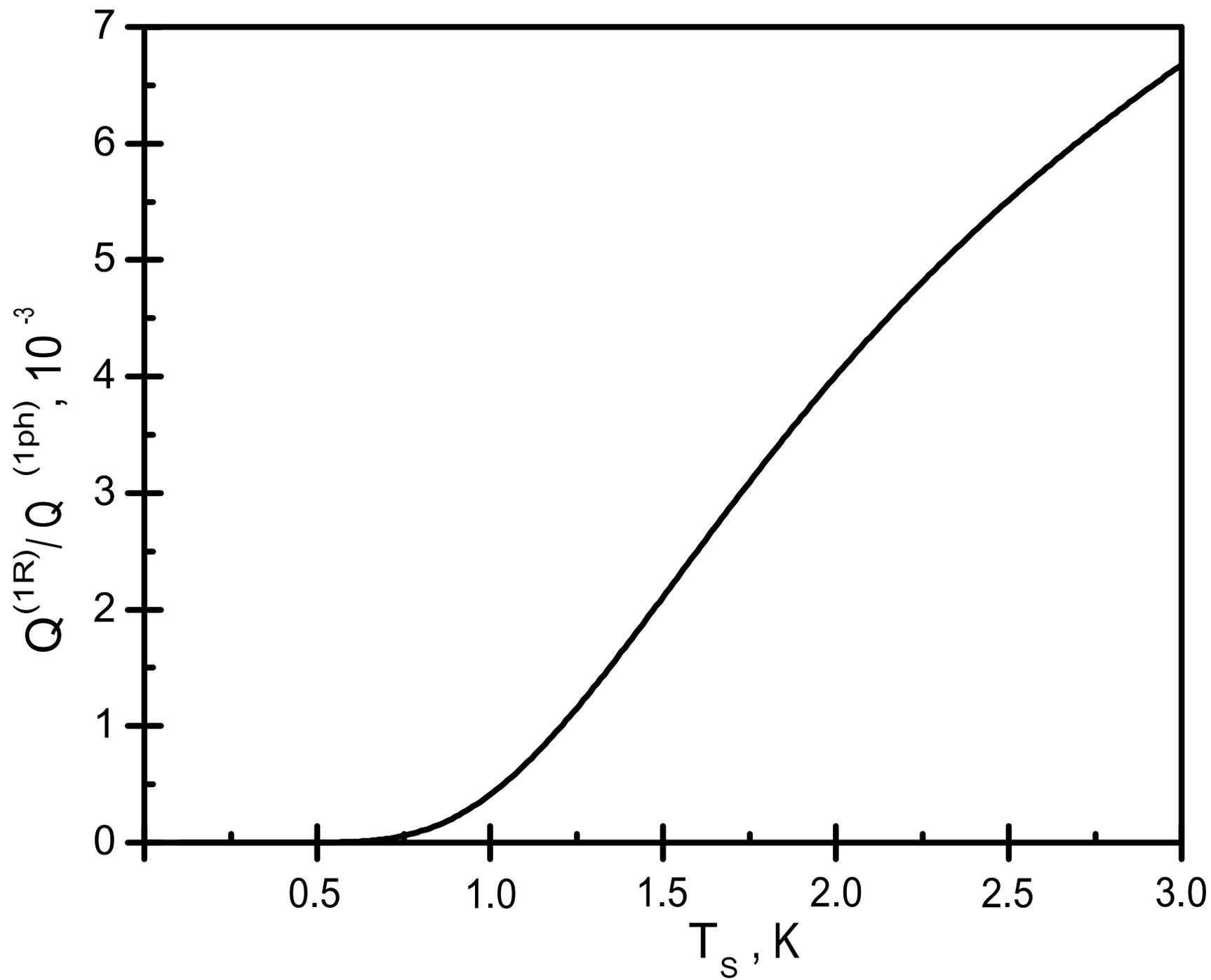
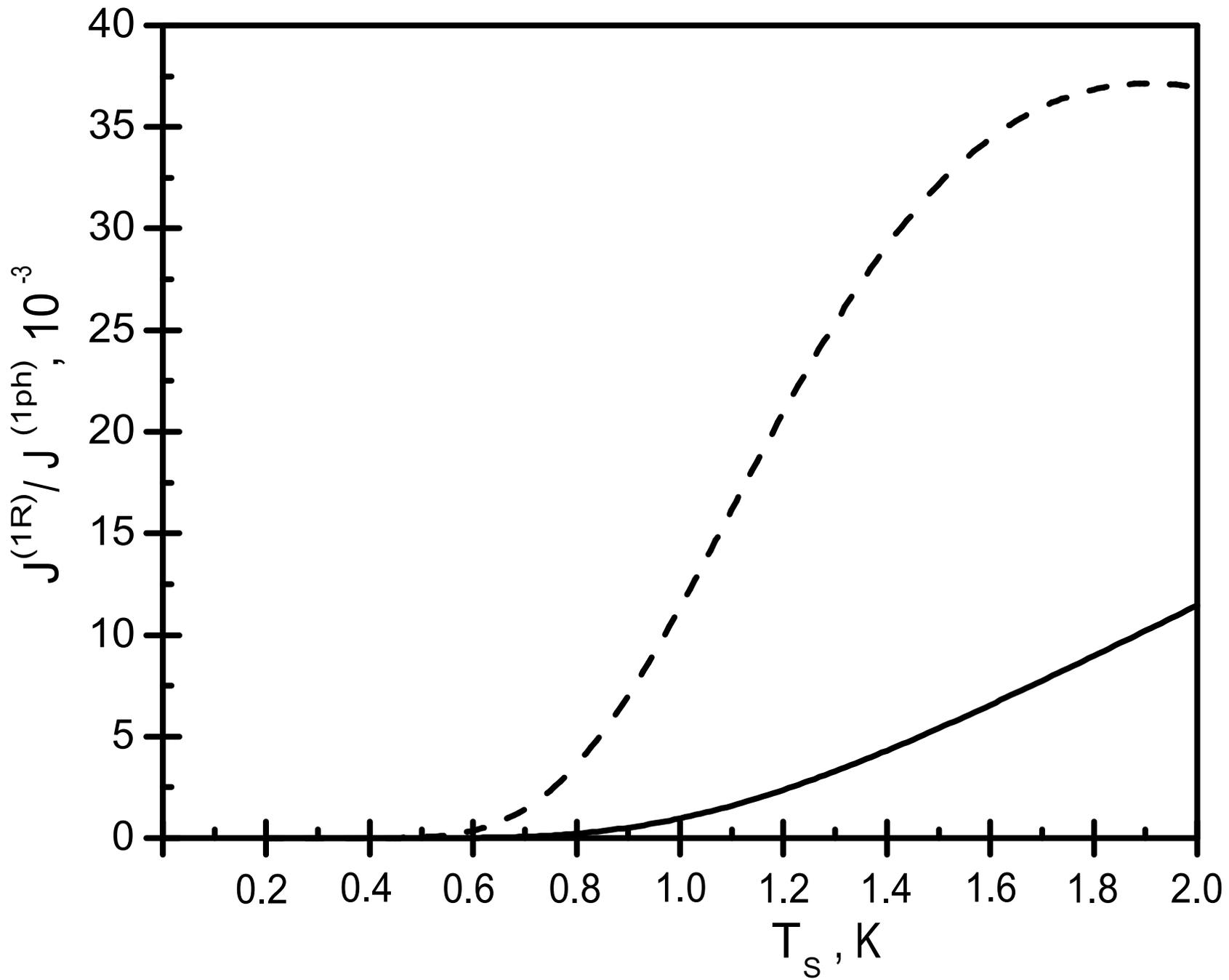
\end{document}